\documentclass[11pt]{article}

\newcommand{\mb}{\mathbf}
\newcommand{\bs}{\boldsymbol}

\RequirePackage{amsthm,amsmath,amsfonts,amssymb}
\usepackage{natbib}
\usepackage{graphicx}

\usepackage{soul}
\usepackage{natbib}
\setcitestyle{authoryear,open={(},close={)}}

\usepackage{hyperref}
\usepackage{color}
\usepackage{multirow}

\usepackage{upgreek}
\usepackage{algorithm2e}
\usepackage{makecell}
\usepackage{multirow}
\usepackage{adjustbox}
\usepackage{url}
\usepackage{setspace}
\usepackage{graphicx}
\usepackage{amsmath}
\usepackage{bbm}
\usepackage{comment}
\usepackage{setspace}
\newcommand{\blind}{1}

\begin{document}
\def\spacingset#1{\renewcommand{\baselinestretch}%
	{#1}\small\normalsize} \spacingset{1}

\if1\blind
{
	\title{\bf High Resolution Global Precipitation Downscaling with Latent Gaussian Models and Nonstationary SPDE Structure}
	\author{Jiachen Zhang\\
	Department of Applied and Computational Mathematics and Statistics,\\
		University of Notre Dame (USA)\\
		and \\
		Matthew Bonas\\
		Department of Applied and Computational Mathematics and Statistics,\\
		University of Notre Dame (USA)\\
		and\\
		Diogo Bolster\\
		Department of Civil and Environmental Engineering and Earth Siences\\
University of Notre Dame (USA)\\
		and\\
		Geir-Arne Fuglstad\\
		Department of Mathematical Sciences\\ Norwegian University of Science and Technology (NTNU)\\ Trondheim, Norway
		\\and\\
		Stefano Castruccio\thanks{scastruc@nd.edu}\hspace{.2cm}\\
		Department of Applied and Computational Mathematics and Statistics,\\
		University of Notre Dame (USA)\\
		\\
	}
	\maketitle
} \fi

\newpage

\begin{abstract}
Obtaining high-resolution maps of precipitation data can provide key insights to stakeholders to assess a sustainable access to water resources at urban scale. Mapping a nonstationary, sparse process such as precipitation at very high spatial resolution requires the interpolation of global datasets at the location where ground stations are available with statistical models able to capture complex non-Gaussian global space-time dependence structures. In this work, we propose a new approach based on capturing the spatial dependence of a latent Gaussian process via a locally deformed Stochastic Partial Differential Equation (SPDE) with a buffer allowing for a different spatial structure across land and sea. The finite volume approximation of the SPDE, coupled with Integrated Nested Laplace Approximation ensures feasible Bayesian inference for tens of millions of observations. The simulation studies showcase the improved predictability of the proposed approach against stationary and no-buffer alternatives. The proposed approach is then used to yield high resolution simulations of daily precipitation across the United States.
\end{abstract}

\noindent%
\vfill

\spacingset{2} 

\addtolength{\textheight}{.5in}%

\section{Introduction}
\label{sec:intro}

Accurate high-resolution information of precipitation data is essential to effective prediction and management of water resources \citep{clark2015}. Dramatic improvements in modeling physical processes driving precipitation have resulted in more realistic simulations from global climate models and hence more reliable predictions. The high complexity of modern climate models, however, implies a computational and storage cost which limit the spatial resolution at which global climate simulations can be performed. As such, there are significant uncertainties and mismatches with observations, due to precipitation patterns that coarse resolutions do not sufficiently represent as they cannot capture the scale of the physical processes of interest \citep{wood2021}. The consequences can be over- or under-attribution of a particular location or incorrect timing of events, that can for example be the difference between a local flooding or not \citep{sapountzis2021}. It is therefore of high scientific interest to refine global predictions and produce maps of both probability of rain occurrence and precipitation intensity at a high spatial scale, in order to inform impact assessment models for flood resilience and agricultural models for drought predictions. 

It is in principle possible to produce high resolution precipitation using a coarse global dataset as boundary condition for a regional weather model such as the Weather and Research Forecasting  (WRF, \cite{ska19}). This \textit{dynamical downscaling} approach \citep{sai11} has the appealing advantage of producing physically consistent spatial fields at high resolution, but comes with a substantial associated cost in terms of computational and storage resources, as well as expertise for model setup that only few research centers, universities or businesses could afford. A more affordable solution lies in the formulation of an empirical relationship between global data and ground observations to be fit at locations where ground data are available. Under the assumption that this relationship is at least approximately valid at unobserved locations, high resolution maps can be produced by correcting the global dataset. This \textit{statistical downscaling} approach \citep{ber10} is fast, computationally affordable, and has a long established track record of success in the geoscience literature. In order to work, such approach requires that the global and the ground data are co-located, which is not a priori the case since global data are defined as averages over large areas. It becomes therefore necessary to use spatial statistical models to interpolate the global simulation values at the same locations of the ground observations, and to have an assessment of the uncertainty around these estimates.

Global spatial data require the formulation of specialized models whose theoretical properties are substantially different from spatial processes on Euclidean spaces. In fact, \cite{gne13} highlighted how a valid process on the sphere with great circle distance could be achieved only with severe restrictions on the parameter space of the most widespread covariance model, the Mat\'ern function. In the past two decades, new modeling approaches tailored for global data have emerged. Among them, \cite{jun07,jun08} proposed to embed the sphere in a three dimensional space, consider a Mat\'ern model and apply partial derivatives to achieve more flexibility. The proposed class of models was able to capture not just an isotropic behavior, but also \textit{axial symmetry}, i.e., a nonstationary behavior across latitude \citep{jon63}. \cite{jun11} generalized this approach to multivariate global processes. A fast and flexible spectral class of axially symmetric models was proposed in the case of gridded data by \cite{cas13}. The approach was then generalized to non-parametric spectral estimation \citep{cas14}, three-dimensional variables \citep{cas16}, different land/ocean behavior \citep{cas17} and also multivariate processes \citep{edw19}. On the more theoretical side, substantial progress has been made in the determination of properties of high dimensional spheres for isotropic processes via basis decomposition see, e.g., \cite{ara20,por20}. We refer to \cite{jae17,por18} for two recent reviews on the topic. 

A novel, different perspective was raised in the seminal work of \cite{lin11}, where a subclass of Mat\'ern models was associated with the solution of a diffusion-reaction Stochastic Partial Differential Equation (SPDE) with the Markov property and inference was performed with finite volumes. The key insight of this approach, as far as global models are concerned, is that the original SPDE on the plane can be just adapted to the sphere, with the additional benefit of not requiring boundary conditions. While in its original formulation the SPDE resulted in stationary models, non-stationary extensions have been proposed by allowing spatially varying coefficients. Several alternatives have been proposed, from nested SPDE \citep{bol11} to models with physical barriers \citep{bak19}. Recently, \cite{fug15,fug20} extended this approach by allowing models with local deformation of the SPDE via a spatially varying scalar and vector field. The proposed approach showed promising results, but has been so far limited to the Gaussian case and generalization to non-Gaussian data is by no means straightforward, given the challenges in modeling non-Gaussian data and the computational overhead implied by these models. 


In this work, we propose a non-Gaussian, non-stationary SPDE-based global spatio-temporal model with local deformation and a buffer between land and sea to account for abrupt changes in spatial dependence. Non-Gaussianity is modeled via a latent Gaussian model, i.e., by assuming that the non-Gaussian marginal behavior is conditionally independent across locations, and then the spatial dependence is captured via a latent process with a Gaussian structure. Inference is still achievable for very large datasets by means of 1) a sparse precision matrix of the latent Gaussian model emerging from the finite volume solution of the SPDE and 2) a fast approximation of the high-dimensional integrals required for posterior computation via Integrated Nested Laplace Approximation (INLA, \cite{rue09}). The model is ideally suited to highly non-Gaussian data such as daily global precipitation, and it is then used to 1) fit global reanalysis data, 2) provide interpolated data at the same location as the ground observations, 3) downscale precipitation using both ground and interpolated data, so that 4) high resolution maps of precipitation are provided. 

The work proceeds as follows. Section \ref{sec:data} introduces the data which will be used in this work. Section \ref{sec:method} details the methodology for the latent Gaussian model, specifically the temporal and the spatial component. Section \ref{sec:inference} shows how inference is performed and how sparsity and numerical approximations alleviate the computational burden. Section \ref{sec:sim} assesses numerically the posterior consistency, as well as the improved predictability of the proposed model against simpler alternatives. Section \ref{sec:app} applies the proposed model to the precipitation data and shows it can provide high resolution maps of daily precipitation across the continental United States. Section \ref{sec:conc} concludes with a discussion. For reproducibility, at the end of this work we provide information about the repository where the code and data are available.

\section{Data Description}
\label{sec:data}
\begin{figure}[!tb]
\centering
\includegraphics[width = 14cm]{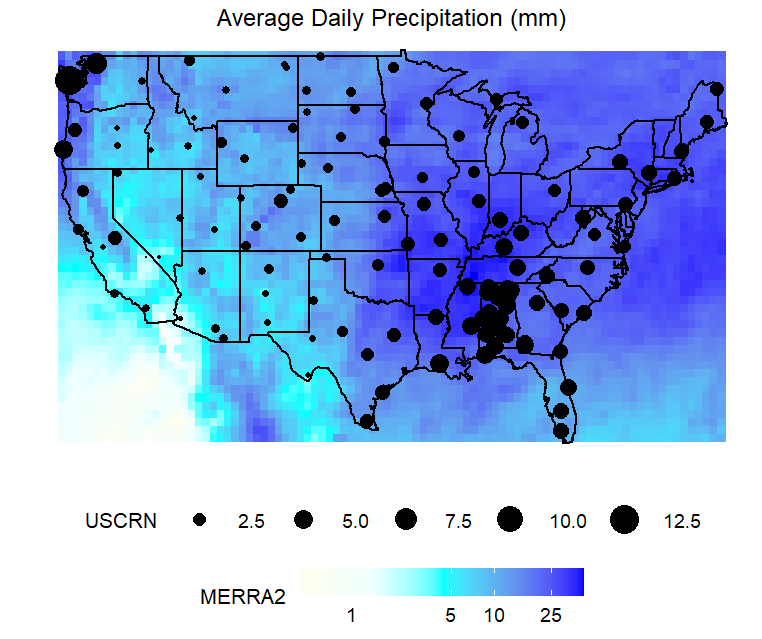}
\caption{Average daily precipitation (in mm) for each USCRN site and MERRA-2 grid point from January 1$^{\text{st}}$, 2021 through December 31$^{\text{th}}$, 2021.}
\label{fig:USCRNLocs}
\end{figure}

\quad We focus on daily global precipitation data from the Modern-Era Retrospective Analysis for Research and Applications, version 2 (MERRA-2, \cite{gel:17}) produced by the NASA Global Modeling and Assimilation Office (GMAO). MERRA-2 is a reanalysis data product that incorporates observations from satellite instruments and is considered one of the best representations of the state of the Earth's system. The data is available on a regular grid with a resolution of $0.625^{\circ}\times 0.5^{\circ}$ in longitude and latitude, respectively, for a total of $n=207,936$ locations. We focus on the year 2021, the latest year with a continuous record available, and we use the daily Maximum Rainfall Rate (MRR, in $\text{kg/}\text{m}^2\cdot \text{s})$. To convert the MRR into precipitation, we divided it by the water density, 1,000 $\text{(kg/m}^3)$, and convert the unit to millimeter by multiplying by 1,000, as well as multiply by 86,400$s$ to obtain the daily precipitation. We assume that for each location, the MRR lasts for the whole day, which leads to some overestimation, as it can be clearly seen from the two different legend scales in Figure \ref{fig:USCRNLocs}. The downscaling approach in Section \ref{sec:app} will be able to account for this by performing a linear transformation between (interpolated) MERRA-2 and USCRN.

For ground observations, we consider the U.S. Surface Climate Reference Network (USCRN, \cite{noaa}), a data product containing continuous records from climate monitoring stations across the continental United States. The USCRN monitoring stations record measurements for total precipitation, measured in millimeters (mm), in real-time in 5-minute intervals. The data are collected with a Geonor T-200B precipitation gauge, whose maximum capacity is 600mm. This gauge uses a precipitation collection bucket which is surrounded by a wind/snow shield and heated in order to prevent ice buildup in cold regions. Three wires attached to this collection device vibrate with frequencies relative to the weight of the bucket, and these vibration frequencies are then converted to gauge depth (in mm). For this work, we consider data from 131 different monitoring stations post-processed to daily resolution forming a continuous record from January $1^{\text{st}}$, 2021 to December, 31$^{\text{th}}$ 2021. Figure \ref{fig:USCRNLocs} shows the locations of the USCRN sensors along with the average total daily precipitation throughout 2021. For comparison, the same figure also shows the average daily precipitation for the MERRA-2 grid points during the same time frame. It is readily apparent from this Figure that the regions of highest average daily precipitation are the northwest and southeast regions of the country whereas the drier region of the country spans from the eastern border of California through to the Mississippi River. 

\section{Methodology}\label{sec:method}

\subsection{Latent Gaussian Model}

\quad We propose a spatio-temporal latent Gaussian model \citep{rue09}, defined for a generic spatial point on the sphere $\mathbf{s} \in \mathbb{S}^2$  and time $t=1, 2, \ldots$ as:
\begin{subequations} \label{eq:latent}
\begin{flalign}
& \qquad Y(\mb{s},t) \mid \mu(\mb{s},t), \bs{\theta}_{\text{MRG}} \sim h (\mu(\mb{s},t),\bs{\theta}_{\text{MRG}}),\label{eqn:latent1}\\
& \qquad g(\mu(\mb{s},t)) = \sum_{p=1}^P\beta_pf_p(\mb{s})+f^{\text{time}}(\mb{s},t)+\epsilon(\mb{s}),\label{eqn:latent2}\\
& \qquad f^{\text{time}}(\mb{s},t) = \sum_{k=1}^K \left\{\zeta_{k}(\mb{s}) \sin\left(\frac{2\pi k t}{\delta}\right)+\zeta'_{k}(\mb{s}) \cos\left(\frac{2\pi k t}{\delta}\right)\right\},\label{eqn:latent3}
 \end{flalign}
\end{subequations}

\noindent where $h(\cdot)$ represents the marginal distribution of $Y(\cdot)$ conditional on the latent field and the hyperparameters, and belongs to the exponential family with some mean $\mu(\mb{s},t)$, whose structure is determined by a latent Gaussian process through a link function $g(\cdot)$. The marginal parameters $\bs{\theta}_{\text{MRG}}$ characterize moments higher than the first, and could be empty. If the marginal distribution is Gaussian, we have $Y(\mb{s})\sim \mathcal{N}(\mu(\mb{s},t),\bs{\theta}_{\text{MRG}})$, and the link function $g(\cdot)$ is simply the identity function \citep{dun18}. For example, if the marginal distribution is the Bernoulli distribution instead, we have $Y(\mb{s})\sim \mathcal{B}(\mu(\mb{s},t))$, and the logit function can be chosen as the link function \citep{dun18}. We assume that the transformed mean in the latent space $g(\mu(\mb{s},t))$ is modeled by a location specific time effect, $f^{\text{time}}(\mb{s},t)$, $p=1, \ldots, P$ location-specific covariates $f_p(\mb{s})$, and a spatial error $\epsilon(\mb{s})$. The time effect $f^{\text{time}}(\mb{s},t)$ is described by $K$ harmonics with parameters $\bs{\zeta}(\mb{s})=(\zeta_1(\mb{s}), \ldots, \zeta_K(\mb{s}))^\top$ and $\bs{\zeta}'(\mb{s})=(\zeta'_1(\mb{s}), \ldots, \zeta'_K(\mb{s}))^\top$. If we assume that we have a sample observed at $\mb{s}_1, \ldots, \mb{s}_n$, the total number of temporal parameters in equation \eqref{eqn:latent3} is $\bs{\theta}_{\text{time}}=\{\bs{\theta}_{\text{time}}(\mb{s}_1), \ldots, \bs{\theta}_{\text{time}}(\mb{s}_n)\}$, where $\bs{\theta}_{\text{time}}(\mb{s}_i)=\{{\bs{\zeta}(\mb{s}_i), \bs{\zeta}'}(\mb{s}_i)\}$, for a total of $2Kn$ parameters. The period $\delta\in \{365, 366\}$ depends on the leap/no-leap year considered. We assume that the spatial random effect $\epsilon(\mb{s})$ is a realization from a mean-zero Gaussian random field independent in time, whose covariance function depends on some parameters $\bs{\theta_{\text{space}}}$ which will be specified in the next Section.

\subsection{Spatial Correlation Structure}
\quad The simplest models for the spatial dependence of $\epsilon(\mb{s})$ are stationary and isotropic, i.e., they assume that the dependence is a function of $\|\mb{s}_1-\mb{s}_2\|$. Among them, one of the most popular choices is arguably the Mat\'ern model, whose correlation between two locations $\mb{s}_1, \mb{s}_2$ is defined as \citep{ste99}

\[
    \text{Corr}(\epsilon(\mb{s}_1),\epsilon(\mb{s}_2))=C(\mb{s}_1,\mb{s}_2)=\frac{1}{2^{\nu-1}\Gamma(\nu)}\left(\frac{\|\mb{s}_1-\mb{s}_2\|}{\rho}\right)^{\nu}K_{\nu}\left(\frac{\|\mb{s}_1-\mb{s}_2\|}{\rho}\right),
\]

where $K_\nu$ is the modified Bessel function of the second kind with smoothness parameter $\nu>0$ (i.e., controlling the degree of mean squared differentiability) and range parameter $\rho>0$. If inference is sought for a large dataset, a matrix comprising of the covariance among all locations could not be stored, and likelihood evaluation could become computationally challenging or just impossible. Instead of operating directly with the covariance matrix, a popular solution in the past decade has been to rely on the identification of a Gaussian process with Mat\'ern covariance as the (unique) stationary solution of the following fractional reaction diffusion SPDE \citep{whi54}:
\begin{equation}\label{eq:spde}
    \left(\frac{1}{\rho^2}-\Delta\right)^{\nu/2+1/2}\epsilon(\mb{s})=\mathcal{W}(\mb{s}),\; \mb{s}\in \mathbb{R} ^2,
\end{equation}
where $\Delta$ is the Laplacian operator and $\mathcal{W}(\mb{s})$ is a spatial Gaussian white noise. By exploiting an `explicit link' between a continuous Markov process when $\nu$ is integer in \eqref{eq:spde} and a discrete Gaussian Markov Random Field (GMRF), \cite{lin11} proved that if all locations are arranged on a 2D lattice, then the covariance structure of the GMRF could be approximated by applying the convolution of a sparse precision matrix. Moreover, any location that is not on the lattice could also be interpolated and approximated by means of a triangulation over the domain. Ultimately, this implies that the Mat\'ern covariance can be approximated by a sparse precision matrix, and hence allow faster and feasible inference on the spatial structure of $\epsilon(\cdot)$. In this work, we rely on a similar SPDE defined on a sphere defined as 
\begin{equation}\label{eq:spdeshpere}
    \left(\frac{1}{\rho^2}-\Delta_{\mathbb{S}^2}\right)^{\nu/2+1/2}\epsilon(\mb{s})=\mathcal{W}(\mb{s}),\;\mb{s}\in \mathbb{S} ^2,
\end{equation}
where $\Delta_{\mathbb{S}^2}$ is the Laplacian operator. 

The aforementioned SPDE approach has clear computational advantages, but in its formulation is limited to stationary and isotropic processes \citep{rue09}. The SPDE operator can, however, be generalized to allow for nonstationary constructs, while still yielding sparse precision matrices. In this work we rely on a spatially varying SPDE originally formulated in \cite{fug19}, but other approaches for spatially varying parameters \citep{rue09} or nested SPDE \citep{bol11} have been proposed. We assume a location on the sphere has polar coordinates $\mb{s}=(L,l)$, where $L$ is the latitude and $l$ is the longitude. We introduce two terms: a vector field $\mb{v}(\cdot)=(v_1(\cdot),v_2(\cdot))^\top$ and a positive-valued scalar field $\rho(\cdot)$. We then define the inverse deformation tensor as:

\[
    \mb{G}(\mb{s})^{-1}=\rho(\mb{s})^2\frac{\mb{I}_2+\mb{v}(\mb{s})\mb{v}(\mb{s})^\top}{\sqrt{1+\|\mb{v}(\mb{s})\|^2}}.
\]

One can show that with the spatially varying metric tensor defined above, the distance along the direction $\mb{v}(\mb{s})$ is scaled by $1/(\rho(\mb{s})(1+\|\mb{v}(\mb{s})\|^2)^{\frac{1}{4}}$. In the orthogonal direction of $\mb{v}(\mb{s})$, the distance is scaled by $(1+\|\mb{v}(\mb{s})\|^2)^{\frac{1}{4}}/{\rho(\mb{s})}$. Therefore, the vector field $\mb{v}(\cdot)$ specifies the direction of the local anisotropic effect at each location, while $\rho(\cdot)$ represents its strength. After specifying the metric tensor $\mb{G}(\mb{s})$, it case be shown that an appropriate change of variable in the SPDE \eqref{eq:spdeshpere} yields \citep{fug20}:

\begin{equation}
    \label{EQ:Gs}
    [|\mb{G}(\mb{s})|^{\frac{1}{2}}-\nabla\cdot|\mb{G}(\mb{s})|^{\frac{1}{2}}\mb{G}(\mb{s})^{-1}\nabla]\epsilon(\mb{s})=|\mb{G}(\mb{s})|^{\frac{1}{4}}\mathcal{W}(\mb{s}),\;\mb{s}\in \mathbb{S} ^2.
\end{equation}

\subsection{Spherical Harmonics}
\quad Both the vector field $\mb{v}(\cdot)$ and the scalar field $\rho(\cdot)$ can be specified through basis decomposition such as spherical vector harmonics and spherical harmonics, respectively. However, a more flexible approach is necessary for global models, which must account not just for slowly changing nonstationarity, but also for abrupt changes dictated by large geographical descriptors such as land and ocean \citep{cas17}. In order to formulate a valid model via SPDE while still accounting for abrupt changes, we consider the buffering approach proposed by \cite{bak19}. More specifically, we use a buffer area along coastlines with a separate parameter that describes the multiplicative drop $d\in [0,1]$ in the strength of dependence in the buffer area, so that for each of the land/ocean domain we propose a separate spherical harmonics decomposition:
\[
    \text{log}\{\rho^j(\mb{s})\}=\sum_{l=0}^\mathcal{L}\sum_{m=-l}^l\alpha_{ml}^jY_l^m(\mb{s}),
\]

\noindent where $\alpha_{ml}^j$ are real-valued coefficients and $Y_l^m(\mb{s})$ are Laplace's spherical harmonic of degree $l$ and order $m$ and $j=\{\text{land, ocean}\}$ specifies the geographical descriptor where $\mb{s}$ is located. Similarly, the vector field $\mb{v}(\cdot)$ can be described as:
\[
    \mb{v}^j(\mb{s})= \sum_{l=1}^\mathcal{L}\sum_{m=-l}^l\{E_{lm}^{(1,j)}\nabla Y_m^l(\mb{s})+E_{lm}^{(2,j)}\hat{\mb{r}}(\mb{s})\nabla \times Y_m^l(\mb{s})\},
\]

\begin{sloppypar}
\noindent where $\hat{\mb{r}}$ is the unit vector in the positive radial direction, $E_{lm}^{(1,j)}$ and $E_{lm}^{(2,j)}$ are real coefficients, $\mathcal{L}$ is the highest order in the bases. Additionally, in order to account for micro-scale variability, we assume that the process for both land and sea also has a nugget $\tau_j^2$. In summary, the spatial parameters of the model are $\bs{\theta}_{\text{space}}=\left\{d,\left\{\tau_j^2, j \in \{ \text{land, sea} \}\right\},\left\{\alpha_{ml}^j, E_{lm}^{(1,j)}, E_{lm}^{(2,j)}, m=-l,\ldots, l; l=1, \ldots, \mathcal{L}, j \in \{\text{land, sea}\}\right\}\right\}$, for a total of $6(\mathcal{L}^2+2\mathcal{L})+3$ parameters. 
\end{sloppypar}

\quad We use a priori independent standard normal distributions as priors for all parameters, with log transformation if they are constrained to be positive. The same setting is applied to the parameters used in simulation study and application.

\section{Inference}\label{sec:inference}
\quad We propose a stepwise inference approach to reduce the overall dimension of the parameter space in each step. We first estimate $\bs{\theta}_{\text{time}}$ at each location independently, then $\bs{\theta}_{\text{space}}$ conditionally on the temporal parameters. In \cite{edw20} it was shown that the stepwise approach results in an asymptotically consistent inference, and \cite{cas17} showed that uncertainty and bias propagation have small impact for large yet finite datasets such as the one we work with here.

\subsection{Step 1: Temporal Structure}
\quad In the first step, the inference is performed at each location independently without spatial and covariate effect. We redefine equation \eqref{eq:latent} as the following: 
\begin{equation} \label{eq:temp}
 \begin{array}{rcl}
 Y(\mb{s}, t) & \sim & h (\mu(\mb{s},t),\bs{\theta}_{\text{MRG}}),\\[7pt]
 g(\mu(\mb{s},t)) & = & \sum_{p=1}^P\beta_pf_p(\mb{s})+\sum_{k=1}^K \left\{\zeta_{k}(\mb{s}) \sin\left(\frac{2\pi k t}{\delta}\right)+\zeta'_{k}(\mb{s}) \cos\left(\frac{2\pi k t}{\delta}\right)\right\}.
 \end{array}
\end{equation}

\noindent The vector of temporal parameters $\bs{\theta}_{\text{time}}$ and the linear parameters $\beta_1, \ldots, \beta_p$ are estimated using least-squares and the parameters are considered fixed in the following inference steps. Once $\hat{\bs{\theta}}_{\text{time}}, \hat{\beta}_1, \ldots, \hat{\beta}_p$ are obtained, conditional on them the spatial parameters $\bs{\theta}_{\text{space}}$ of the spatial process $\epsilon(\mb{s})$ can be estimated.

\subsection{Step 2: Spatial Covariance Structure}

\quad We define a collection of triangles $T_1, \ldots, T_{n_T}$ on the sphere, and use a finite volume method to discretize the SPDE in \eqref{EQ:Gs}. We redefine the inverse matrix tensor as $\mb{G}(\mb{s})^{-1}=\rho(\mb{s})^2\mb{H}(\mb{s})$, where $|\mb{H}(\mb{s})|=1$, and we integrate it over triangles $T_i$ generated on a global mesh and seek for a piece-wise constant solution to the SPDE. For all triangles $T_i$, we have the following equality in distribution:
\begin{equation}
    \label{EQ:joint}
    \left[\int_{T_i}\frac{1}{\rho(\mb{s})^2}-\nabla\cdot \mb{H}(\mb{s})\nabla\right]\epsilon(\mb{s})\mathrm{d}V \overset{d}{=} \int_{T_i}\frac{1}{\rho(\mb{s})}\mathcal{W}(\mb{s})\mathrm{d}V.
\end{equation}

\noindent Here $\nabla\cdot$ is the divergence operator, $\nabla$ is the gradient operator, and $\mb{H}(\cdot)$ is a $2\times2$ piecewise continuously differentiable diffusion tensor and $\mathrm{d}V$ is the surface measure on the triangles. This allows to translate the SPDE into a set of linear equations for a Gaussian vector that is assumed to be constant across each triangle.

Similarly to \cite{bert07,fug20}, let $\bs{\epsilon}=(\epsilon_1,\epsilon_2,...,\epsilon_n)$ be the vector of values at triangle center, then the following $n\times n$ matrix $\mb{A_H}$ could be calculated to describe a discrete approximation:
\[
    \left(\sum_{j=1}^3\int_{\sigma_{i,j}}(\mb{H}(\mb{s})\nabla \epsilon(\mb{s}))^\top n_{i,j}\mathrm{d} \mb{s}\right)_{i=1}^n \approx \mb{A_H}\bs{ \epsilon}.
\]

\noindent Here, $\sigma_{i,j}$ represents the three faces of the triangle $T_i$. Then, we combine this with a $n\times n$ diagonal matrix $\mb{D}$, in which $d_{ii}=|T_i|/\rho(x_i)^2$, so that we have:
\[
    \left(\int_{T_i}\frac{\epsilon(\mb{s})}{\rho(\mb{s})^2}\mathrm{d}\mb{s}-\sum_{j=1}^3\int_{\sigma_{i,j}}(\mb{H}(\mb{s})\nabla u(s))^T\mb{n}_{i,j}\mathrm{d}\mb{s}\right)_{i=1}^n\approx(\mb{D}-\mb{A_H})\bs{\epsilon}.
\]

\noindent With this approximation, the equality in distribution expressed in equation \eqref{EQ:joint} can now be expressed as: 
\[
    (\mb{D}-\mb{A_H})\bs{\epsilon}\sim \mathcal{N}(0,\mb{L}),
\]

\noindent where $\mb{L}$ is a $n\times n$ diagonal matrix with elements $l_{ii}=|T_i|/\rho(x)_i^2$. This implies that $\bs{\epsilon}\sim \mathcal{N}(0,\mb{Q}^{-1})$, and $\mb{Q}$ is a sparse precision matrix defined as:
\[
    \mb{Q}=(\mb{D}-\mb{A_H})^\top \mb{L}^{-1}(\mb{D}-\mb{A_H}).
\]

\noindent Therefore, the finite volume method ensures a sparse precision matrix, which mitigates the computational burden for large global data and boosts the computing speed of the nonstationary model during inference.

\subsection{Inference for Latent Gaussian model}
\quad In order to perform inference on the latent Gaussian Model, in this work we make use of the Nested Laplace Approximation (INLA, \cite{rue09}) a method for Bayesian inference alternative to traditional Markov Chain Monte Carlo (MCMC), which could further ease the computational burden. INLA is a deterministic method for fast approximation of high dimensional integrals which takes advantage of computational properties of models that can be expressed as a latent GMRF. Thus, the INLA approach is used for performing the inference in this study. Under the proposed latent Gaussian Model structure, we have the observed data vector denoted here as $\bs{Y}=(Y(\mb{s}_1), \ldots, Y(\mb{s}_n))^\top$ at locations $\mb{s}_i$ that can be described by hyperparameter vector $\bs{\theta}_{\text{space}}$. For simplicity, throughout this section, we will use $\bs{\theta}$ to represent hyperparameter vector $\bs{\theta}_{\text{space}}$. If conditioned on latent spatial field $\mb{X}$, the observations are marginally independent with likelihood: 
\[
    \pi(\mb{Y}|\mb{X},\bs{\theta})=\prod_{i=1} ^n\pi(Y(\mb{s}_i)|X(\mb{s}_i),\bs{\theta}),
\]

\noindent where $\bs{X}=(X(\mb{s}_1),\ldots, X(\mb{s}_n))^\top$ is a Gaussian field with mean zero and modeled by a SPDE approach with precision matrix $\mb{Q}(\bs{\theta})$. Therefore, the joint distribution of latent effect and hyperparameters can be written as:
\[
\begin{array}{rcl}
    \pi(\mb{X},\bs{\theta}|\mb{Y}) &\propto& \pi(\bs{\theta})\pi(\mb{X}|\bs{\theta})\prod_{i=1}^n  \pi(Y(\mb{s}_i)|X(\mb{s}_i),\bs{\theta})) \\[7pt]
    &\propto&
    \pi(\bs{\theta})|Q(\bs{\theta})|^{1/2}\text{exp}\{-\frac{1}{2}\mb{X}^\top Q(\bs{\theta}) \mb{X}\}\prod_{i=1}^n \pi(Y(\mb{s}_i)|X(\mb{s}_i),\bs{\theta}),
\end{array}
\]

\noindent where $|\bs{Q}(\bs{\theta})|$ is the determinant of the precision matrix. The main goal is to approximate the posterior marginals $\pi(X(\mb{s}_i)|\mb{Y})$, $\pi(\bs{\theta}|\mb{Y})$ and $\pi(\theta_j|\mb{Y})$. The marginal posterior distributions of interest can be written as:
\[
\begin{array}{rcl}
\pi(X(\mb{s}_i)|\bs{Y}) &=& \int \pi(X(\mb{s}_i)|\bs{\theta},\bs{Y})\pi(\bs{\theta}|\bs{Y})\mathrm{d}\bs{\theta}\\[7pt]
\pi(\theta_j|\bs{Y})&=&\int \pi(\bs{\theta}|\bs{Y})\mathrm{d}\theta_{-j}.
 \end{array}
\]

\noindent The key idea of INLA approach is to use the form above to construct nested approximations. The approximations of the marginals for the latent field $\pi(X(\mb{s}_i)|\mb{Y})$ are computed by approximating $\pi(\bs{\theta}|\mb{Y})$ and $\pi(X(\mb{s}_i)|\bs{\theta},\mb{Y})$, and using numerical integration to integrate out $\bs{\theta}$. In other words, the posterior marginals of the latent parameter would be obtained by:
\[
\tilde{\pi}(X(\mb{s}_i)|\bs{Y})=\sum_{k}\tilde{\pi}(X(\mb{s}_i)|\bs{\theta}_k,\bs{y})\times \tilde{\pi}(\bs{\theta}_k|\bs{Y})\times \Delta_k,
\]

\noindent where $\Delta_k$ are the weights associated with a vector $\bs{\theta}_k$ of hyperparameters in a grid.

\section{Simulation Studies}\label{sec:sim}

\quad Throughout this section, we denote with NS-LS the proposed nonstationary latent Gaussian model \eqref{EQ:Gs} with land/sea effect with NS the nonstationary model with no land/sea effect. We further consider the stationary SPDE model \eqref{eq:spdeshpere}, and denote with S-LS the model with land/sea effect and with S without it. In Section \ref{sec:hyper}, we perform simulations from the Gaussian marginal distribution for NS-LS to numerically assess posterior consistency for both the hyperparameters and the resulting covariance matrix. In Section \ref{sec:gaus} and Section \ref{sec:bern}, we perform simulations from Gaussian and Bernoulli marginal distributions with identity and logit link, respectively, to assess the interpolation (kriging) performance of the NS-LS against NS, S-LS and S. 

Since the key contribution of this work lies in the spatial component of the model, throughout this section we will assume a purely spatial process with no covariates. In other words, model \eqref{eq:latent} simplifies to 

\begin{subequations} \label{eq:latent:simp}
\begin{flalign}
& \qquad Y(\mb{s}) \sim h (\mu(\mb{s}),\bs{\theta}_{\text{MRG}}),\label{eqn:latent1:simp}\\
& \qquad g(\mu(\mb{s})) = \epsilon(\mb{s})\sim \mathcal{N}(0, \bs{\Sigma}(\bs{\theta_{\text{space}}})).\label{eqn:latent2:simp}
 \end{flalign}
\end{subequations}

In the Gaussian case we also have $\bs{\theta}_{\text{MRG}}=\sigma^2=0.05$, while in the Bernoulli case no marginal parameters are defined, so that $\bs{\theta}_{\text{MRG}}=\emptyset$. 

For each simulation, we sample $n=2,000$ data points on the unit sphere, and then draw the parameters of $\bs{\theta_{\text{space}}}$ from a Normal distribution with mean 1 and standard deviation 0.5, assume them fixed (similar results have been observed for other samples or distributions). Each simulation comprises of $n_r=100$ replicates from the resulting covariance matrix $\bs{\Sigma}(\bs{\theta_{\text{space}}})$. We simulate data from a NS-LS model with $\mathcal{L}=1$, so that there is a total of $6(\mathcal{L}^2+\mathcal{L})+3=21$ hyperparameters. We perform $n_s=100$ independent simulations and report the results both in terms of aggregated performance and their uncertainty .

\subsection{Posterior consistency in the Gaussian case}\label{sec:hyper}

\quad In order to numerically assess posterior consistency, for each simulation we consider an increasing number of replicates $n_r=10, \ldots, 100$. Inference is performed assuming the same model \eqref{eq:latent:simp} and with a mesh of $n_T=2,000$ triangles. For varying levels of $n_r$, the hyperparameters' posterior distributions is retrieved and is compared with the true value. Posterior consistency can be empirically verified in the extent to which the hyperparameters' posterior distributions converges to the true parameters $\bs{\theta_{\text{space}}}$ as $n_r$ increases. 


\begin{figure}[ht!]
\centering
\includegraphics[scale=0.45]{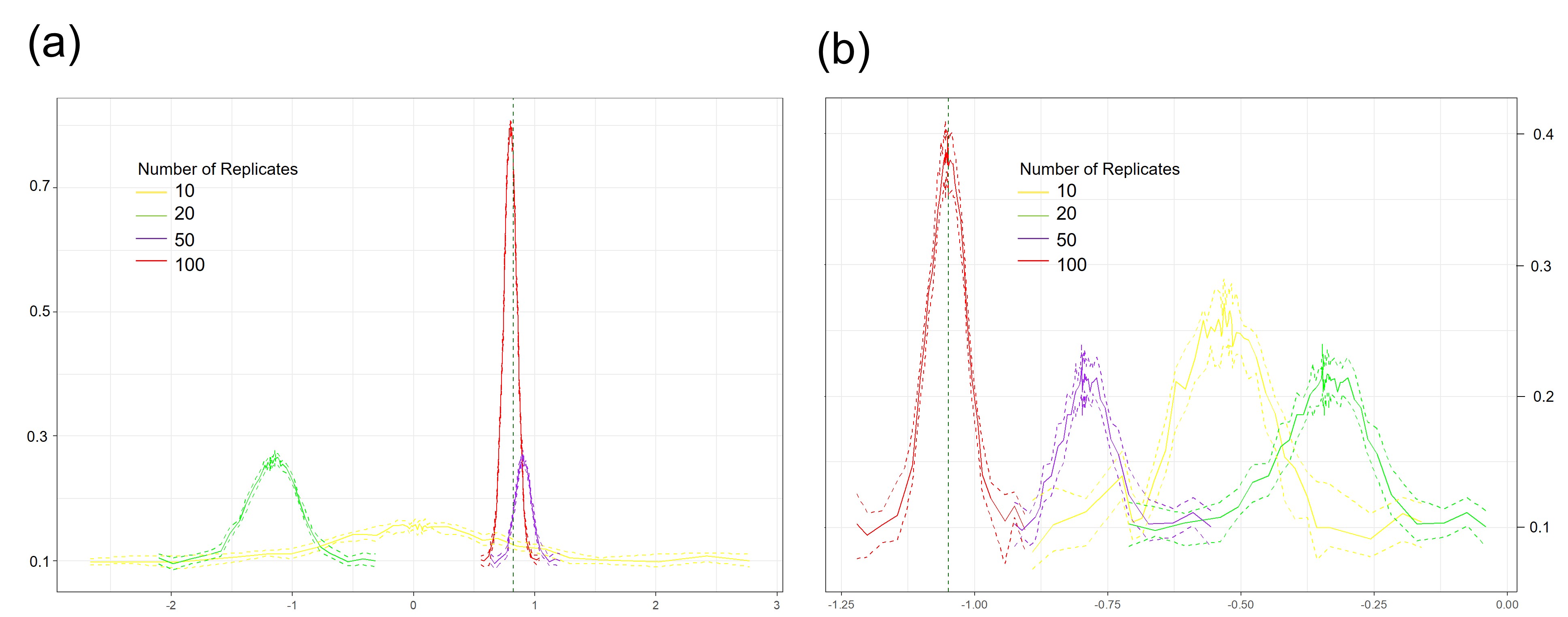}
\caption{Functional boxplots \citep{sun11} across $n_s$ simulations of the posterior distribution of two hyperparameters (a) $\alpha_{11}^2$ and (b) $E_{10}^{(2,2)}$ for different number of replicates $n_r$. The vertical dashed lines represent the true hyperparameter values.}
\label{fig:hyper}
\end{figure} 

\begin{table}[ht!]
	\caption{Median MSE (IQR) between the true hyperparameter and the posterior distribution across all simulations $n_s$ for Gaussian case. \label{tab:hypermse}}
	\centering
	\begin{tabular}{|l|l|l|l|l|l|}\hline 
		$n_r$ & 20 & 40 & 60 &  80 & 100 \\ \hline
		Median MSE (IQR) & 0.32 (0.13) & 0.25 (0.07) & 0.14 (0.05) & 0.05 (0.05) & 0.01 (0.007)\\ \hline
 \end{tabular}
\end{table}

\noindent Figure \ref{fig:hyper} shows the functional boxplot \citep{sun11} for all $n_s$ of the posterior distributions, for two hyperparameters for increasing values of realizations $n_r$. It is readily apparent how the posterior mean aligns to the true parameter value and the posterior standard deviations decreases as the replicates increase. While results are shown for NS-LS, similar patterns have been observed across all other models (NS, S-LS and S). Table \ref{tab:hypermse} shows the median MSE and InterQuartile Range (IQR) of the hyperparameters posterior means estimated from the NS-LS model and the true values across all hyperparameters and across all $n_s=100$ simulations. The median MSE decreases as the replicates increases. 

In order to perform a uniform comparison across all hyperparameters, whose number quickly becomes overbearing (e.g., with $\mathcal{L}=4$ we would have $6(4^2+4)+3=123$ hyperparameters), we also compare the covariance matrix implied by the hyperparameters with the true one. We assess the discrepancy in the covariances via the Kullback-Leibler Divergence (KLD), which in the case of an $n$-dimensional Gaussian distributions with mean $\bs{\mu}_0$ and $\bs{\mu}_1$ and covariance matrices $\bs{\Sigma}_0$ and $\bs{\Sigma}_1$ simplifies to:
\[
    \frac{1}{2}\left(\text{tr}(\bs{\Sigma}_1^{-1}\bs{\Sigma}_0)-n+(\bs{\mu}_1-\bs{\mu}_0)^\top \bs{\Sigma}_1^{-1}(\bs{\mu}_0-\bs{\mu}_1)+\text{ln}\left(\frac{\text{det}\bs{\Sigma}_1}{\text{det}\bs{\Sigma}_0}\right)\right).
\]

\begin{figure}[ht!]
\centering
\includegraphics[scale=0.5]{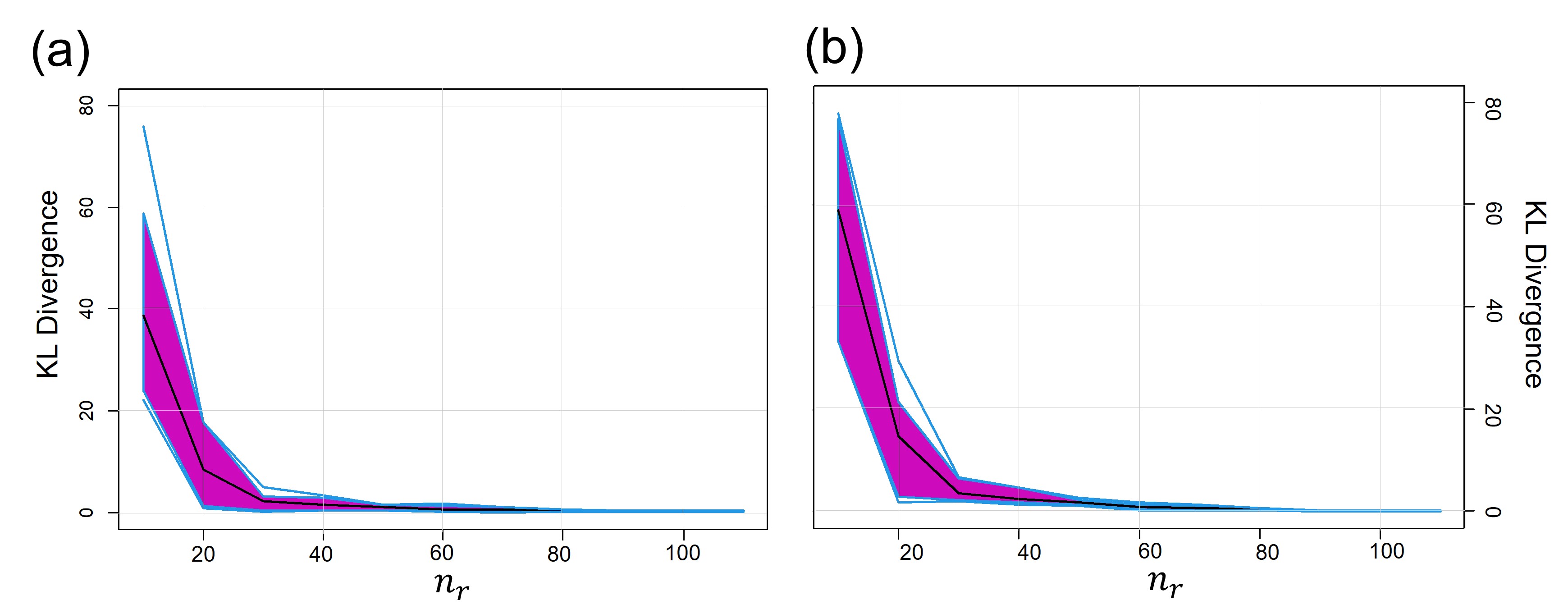}
\caption{Functional boxplot across $n_s=100$ simulations of the KLD between the true covariance matrix and the estimated one according to (a) NS-LS and (b) S-LS.}
\label{fig:KLD}
\end{figure}

\noindent In our case $\bs{\mu}_0=\bs{\mu}_1=\mb{0}$, $\bs{\Sigma}_0=\bs{\Sigma}(\bs{\theta}_{\text{space}})$ and $\bs{\Sigma}_1=\bs{\Sigma}(\hat{\bs{\theta}}_{\text{space}})$, so that the KLD measures the distance between the true and estimated covariance. The results as shown in Figure \ref{fig:KLD} for NS-LS (panel (a)) and S (panel (b)), where the functional boxplot \citep{sun11} of KLD across all $n_s=100$ simulations for an increasing number of realizations $n_r$ is shown. The functional boxplot is used to report the envelope of the 50\% central region (pink area), the median curve (black line) and the maximum non-outlying envelope (outer blue line). As in the case of the estimated parameters, we observe how even with a relatively small number of replicates in the training set, the estimated covariance is converging to the true one. In particular, after 40 replicates the estimated covariance is practically indistinguishable from the true one. 

\subsection{Interpolation performance in the Gaussian case}\label{sec:gaus}
\quad In order to assess the interpolation performance, we perform inference on the hyperparameters for all four models and use them to interpolate at specified locations. We consider two cases (1) all $n$ data points are used in the training set and interpolation is performed at the same sites (2) 92\% of the $n$ locations are considered in the training set, and the others 8\% are withheld for crossvalidation. The test locations are located in within three selected areas indicated in Figure S1. Interpolation performance is measured with the MSE. 

Results for both cases are reported in Table \ref{tab:mse}, and it is readily apparent how the MSE of NS-LS model is the smallest among all four models for the both the all location case (1) and the cross-validation setting (2). More specifically, compared to the S-LS model, the NS-LS model shows an improvement of the median MSE across all locations by 14.6\%. The NS-LS model also shows an appreciable improvement in MSE by 10.4\% and 16.7\%, compared with the NS and S models respectively. From these results it is clear how the land/sea effect and buffer area construction yield significant improvement when used in conjunction with the NS model. 

\begin{table}[ht!]
	\caption{Comparison of interpolation performance across models. The first two columns show the median MSE (IQR) across all $n_s=100$ simulations in the Gaussian case for both (1) all locations and (2) for crossvalidation. The last two columns show the median AUC (IQR) for the Bernoulli case across the same two cases. \label{tab:mse}}
	\centering
	\begin{tabular}{|l|l|l|l|l|l|}\hline 
		Model & locations& NS-LS & S-LS & NS & S  \\ \hline
		\multirow{2}{*}{Gaussian} & All locations & 90.12 (8.17) & 105.47 (9.94) & 100.55 (11.25) & 108.25 (11.23) \\ 
		& Crossvalidation & 9.11 (1.04) & 21.88 (1.18) & 19.35 (1.62) & 21.39 (1.59) \\ \hline \hline
		\multirow{2}{*}{Bernoulli} & All locations & 0.824 (0.048) & 0.769 (0.074) & 0.782 (0.051) & 0.753 (0.050) \\ 
		& Crossvalidation & 0.707 (0.072) & 0.676 (0.081) & 0.672 (0.079) & 0.641 (0.079)\\ \hline
 \end{tabular}
\end{table}

\subsection{Interpolation performance in the Bernoulli case}\label{sec:bern}
\begin{figure}[ht!]
\centering
\includegraphics[scale=0.42]{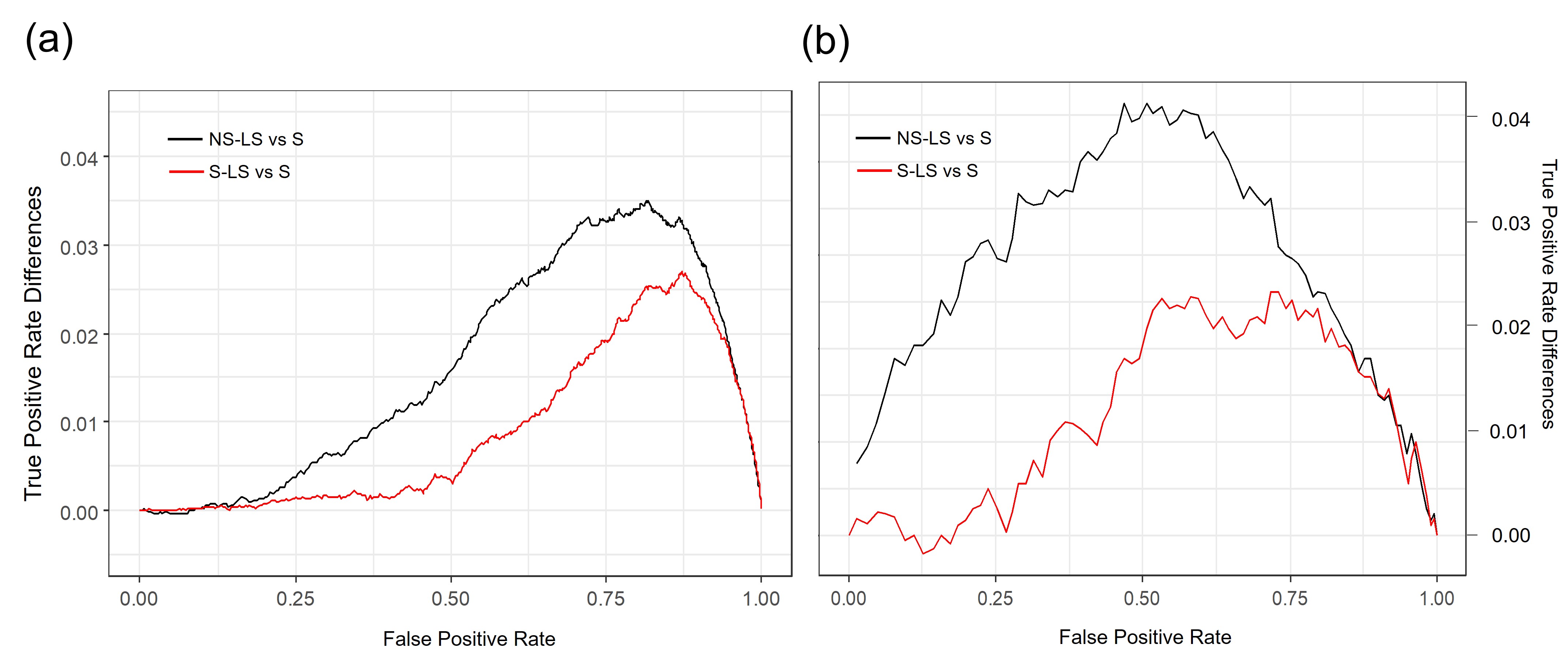}
\caption{Average differences across all $n_s=100$ simulations between ROC curves of NS-LS and S (black line), and S-LS and S (red line) for (a) all locations and (b) cross-validation. The ROC for NS are visually indistinguishable to that of the S-LS model, so the results associated to that model are not show.}
\label{fig:roc}
\end{figure}

\quad We now assess predictability in the case of a Bernoulli distribution with logit link, and as in Section \ref{sec:gaus} we assess both the case where all locations are used as training set, as well as cross-validation with the same testing locations as before. Figure \ref{fig:roc} shows the average differences across all $n_s=100$ simulations between receiver operating characteristic curve (ROC) for NS-LS and S-LS, using S as reference for all locations and validation locations. The ROC for NS are visually indistinguishable to that of the S-LS model, so the results associated to that model are not show. The ROC difference in both cases show how the NS-LS model is uniformly better than the stationary S model (as the ROC difference is always positive), and also uniformly better than the S-LS model, especially in the middle of the curve. As expected, the extent of improvement of NS-LS is larger in the case of cross-validation (panel (b)), where the added value of the model at unobserved locations is more apparent. 

In order to have a comprehensive assessment across all possible choice of thresholds, we consider the area under the curve (AUC) of the ROC for all models and we report it in Table \ref{tab:mse}. In the best case of a perfect prediction, i.e., 100\% true positive rate uniformly across the choice the threshold the AUC should equal 1, and in the worst case of a random guess it should be 0.5. The extent to which the AUC is close to 1 is a measure of predictive performance in this case. As it is shown in Table \ref{tab:mse}, the NS-LS outperforms every other model in both cases. More specifically, across all locations, the NS-LS yields an improvement by 7.2\%, 5.3\% and 9.4\% for the S-LS, NS and S models respectively. These results agree with those presented in Section \ref{sec:gaus}, for the use of the land/sea effect and buffer area construction definitively yields improved performance when included in the NS model.

\section{Application}\label{sec:app}
\quad In this section, we use the data detailed in Section \ref{sec:data} and the proposed latent Gaussian model with nonstationary SPDE introduced in Section \ref{sec:method} to estimate the global probability of a rain event and the precipitation intensity. In section \ref{sec:down}, we discuss both the fit of the global MERRA-2 dataset and the downscaling approach to adjust interpolated MERRA-2 data with ground USCRN precipitation measurement. In section \ref{sec:eval}, we provide evaluation metrics to assess the model performance. 

\begin{figure}
\centering
\includegraphics[scale=0.55]{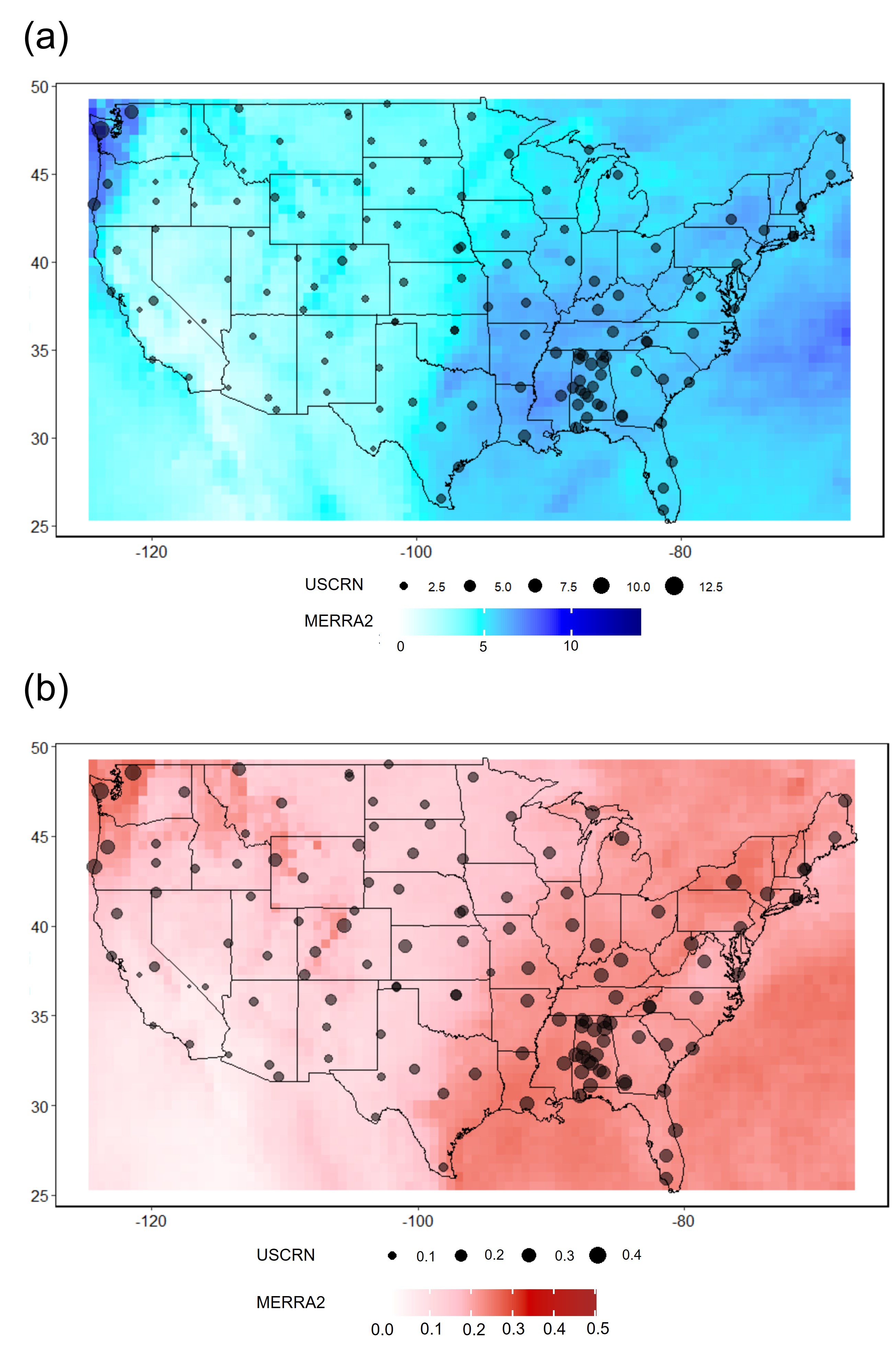}
\caption{Average (a) daily precipitation and (b) precipitation probability. The global dataset is interpolated at the same sites as the ground observations according to the nonstationary global SPDE model \eqref{EQ:Gs}, the linear model \eqref{eq:down} is fit, and the resulting relationship is used to produce the downscaled maps.}
\label{fig:app}
\end{figure}

\begin{figure}[ht!]
\centering
\includegraphics[scale=0.44]{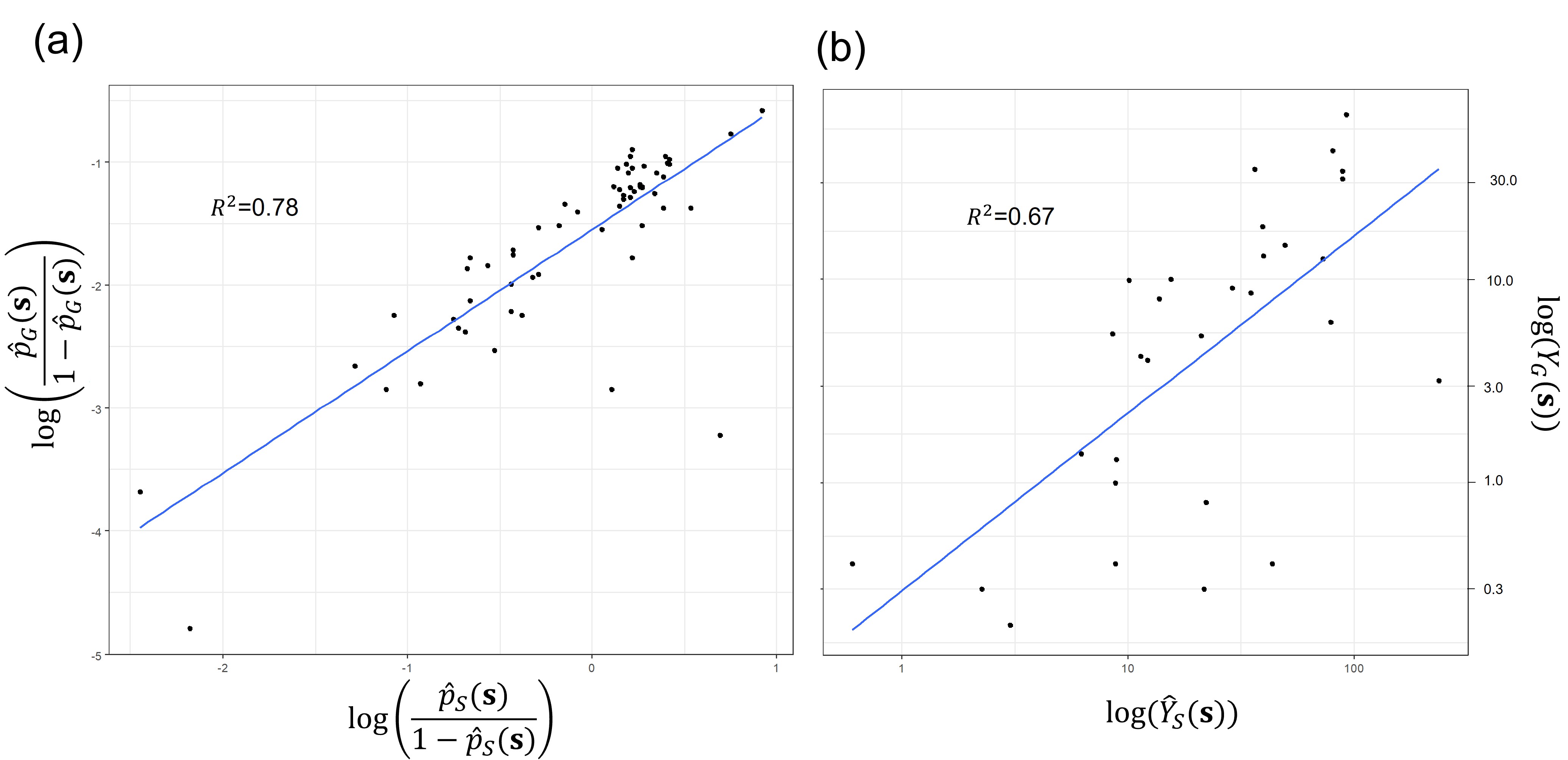}
\caption{The fitted lines using downscaling models described in (a) equation \eqref{eq:down1} and (b) equation \eqref{eq:down2} on February $1^{\text{st}}$, 2021.}
\label{fig:ols}
\end{figure}

\subsection{Modeling global precipitation and downscaling}\label{sec:down}
\quad We initially focus on the MERRA-2 data and consider two global data sets 1) a binary rain occurrence event and 2) in case of rain, the actual rain intensity. We then fit the latent Gaussian model \eqref{eq:latent} with nonstationary PDE \eqref{EQ:Gs} with $\mathcal{L}=1$, using a Bernoulli marginal distribution with a logit link function $g(\cdot)$ for rain occurrence and a Gamma distribution with negative inverse link function for rain intensity. Validation for the choice of the marginal distribution can be found in the supplementary along with Figure S3 showing the histogram of precipitation at 456 sample locations (resolution of $18.75^{\circ}\times 15^{\circ}$ in longitude and latitude) with estimated Gamma density. The sample locations are sparse in space to mitigate any spatial influences. 

In both cases, no additional covariates are assumed, and we assume $K=2$ harmonics for the temporal component, as it was shown to be the optimal choice according to the model selection in Figure S2. Formally, model \eqref{eq:latent} now specializes in the following two models:

\begin{subequations} \label{eq:app}
\begin{flalign}
\log\left(\frac{\mu(\mb{s},t)}{1-\mu(\mb{s},t)}\right) = f^{\text{time}}(\mb{s},t)+\epsilon(\mb{s}),\label{eqn:app1} \quad \text{precipitation probability}\\
 -\mu(\mb{s}, t)^{-1} = f^{\text{time}}(\mb{s},t)+\epsilon(\mb{s}). \quad \text{precipitation intensity} \label{eqn:app2}
\end{flalign}
\end{subequations}

The histogram shows that precipitation intensity follows a Gamma distribution with shape parameter 0.826 and scale parameter 0.184. Inference is performed with a global triangulation of $n_T=2,340$ triangles, of which $1,134$ are within the area of interest (contiguous United States), while the remaining $1,206$ cover the rest of the world. 

The hyperparameters' posterior distributions is obtained and used to predict both the precipitation probability and intensity at the locations where the 131 USCRN ground observations locations are located, see Figure \ref{fig:USCRNLocs}. These predictions are then adjusted (downscaled) to point resolution via linear regression. Since we perform downscaling independently for every time point, for simplicity we now drop the time dependence, and we denote as $Y_{G}(\mb{s})$ and $Y_{S}(\mb{s})$ the precipitation intensity for USCRN and MERRA2, respectively (G=ground, S=simulation), and with $p_{G}(\mb{s})$ and $p_{S}(\mb{s})$ the probability of precipitation occurrence. We further denote as $\hat{Y}_{S}(\mb{s})$ and $\hat{p}_{S}(\mb{s})$ the estimated intensity and probability of occurrence, respectively, according to the proposed SPDE model. Finally, we estimate the probability of precipitation occurrence for the USCRN data by fitting the latent Gaussian model \eqref{eq:latent} for each location independently as a time series model, i.e., assuming no spatial dependence and denote the estimate as $\hat{p}_{G}(\mb{s})$. We further assume a linear relationship between USCRN and MERRA2 precipitation occurrence probability and intensity:

\begin{subequations}\label{eq:down}
\begin{flalign}
\log\left(\frac{\hat{p}_G(\mb{s})}{1-\hat{p}_G(\mb{s})}\right) = \beta_0^{(O)} + \beta_1^{(O)}\log\left(\frac{\hat{p}_S(\mb{s})}{1-\hat{p}_S(\mb{s})}\right)+\xi_O(\mb{s}),\label{eq:down1} \quad \text{precipitation probability}\\
\log\left(Y_G(\mb{s})\right) =\beta_0^{(I)}+\beta_1^{(I)}\log\left(\hat{Y}_S(\mb{s})\right)+\xi_I(\mb{s}),\quad \text{precipitation intensity}\label{eq:down2}
\end{flalign}
\end{subequations}

where $\xi_j(\mb{s})\sim \mathcal{N}(0,\sigma^2_j), j \in \{O,I\}$ independent and identically distributed in space. A functional boxplot of the variogram of the residuals in Figure S4 (with each curve representing a different time point) lends support to the assumption of spatial independence of the error. The downscaling parameters $\beta_0^{(I)}$ and $\beta_1^{(I)}$ for precipitation intensity are then estimated using the ordinary least squares. 

\subsection{Results and Evaluation}\label{sec:eval}

\quad Downscaled probabilities of precipitation occurrence and precipitation intensity according to the aforementioned model are displayed in Figure \ref{fig:app}(a) and (b), respectively, with the dark bubbles representing average values from the USCRN data. The prediction maps of the United States show high daily precipitation and high precipitation intensity around Seattle, while the lowest values can be found near Las Vegas, and overall the model prediction resembles the ground observation values across the United States. To evaluate the model performance, we calculate the root mean squared error (RMSE) for both probability of precipitation occurrence and precipitation intensity. The RMSE for intensity and probability of precipitation occurrence is 2.01 mm and 0.14 mm, respectively. In order to assess the value added by the smoothing of our SPDE model, we also perform downscaling with the linear models in \eqref{eq:down}, but assuming that no spatial model is fit, i.e., that the MERRA-2 data are not interpolated at the locations of the USCRN sites. Instead, we consider MERRA-2 data at their original resolution, and attribute to each USCRN site the value in the same cell. In other words, we consider as covariates $p_S(\mb{s},t)$ and $Y_S(\mb{s},t)$. The resulting RMSE for this model in the case of precipitation intensity and probability of precipitation occurrence is 82.74 mm and 0.28 mm, respectively. Therefore, the proposed SPDE approach has narrowed the discrepancy between MERRA-2 and USCRN significantly, as it has reduced the RMSE for precipitation intensity and probability of precipitation occurrence by 97.6\% and 50\%, respectively. Figure \ref{fig:ols} shows the fitted lines using downscaling model in \eqref{eq:down1} and \eqref{eq:down2} on February $1^{\text{st}}$, 2021. The $R^2$ for the two linear models are 0.78 and 0.67 for precipitation probability and intensity, respectively. 

We also evaluate the model uncertainty by crossvalidation. First, we remove the data from one ground observation location and fit the model using the remaining observations. Next, we construct the 95\% credibility interval for the posterior mean of the probability of precipitation occurrence or precipitation intensity at the removed location with the estimated posterior distributions of the hyperparameters of the model. Then, we repeat the same procedure for all the 131 locations in USCRN. Finally, we determine how many intervals among the 131 the 95\% credibility intervals cover the true value. For precipitation, 93.1\% (122/131) of the 95\% credibility intervals cover the true value, while for probability of raining, 91.6\% (120/131) of the 95\% credibility intervals cover the true value.

\section{Conclusion and Discussion}
\label{sec:conc}

In this work, we have proposed a novel non-stationary spatio-temporal SPDE model able to smooth both probability of precipitation occurrence and probability intensity from a global datasets. Such interpolated dataset is then used in conjuction with ground observation to produce high resolution (downscaled) precipitation maps, which allow to predict what would ground observations would look like in unsampled location with a higher degree of accuracy compared to the original simulated data (i.e., the global data at their native resolution). One may in principle use MERRA-2 as a boundary condition to drive regional simulations with models such as WRF to obtain precipitation maps at equally high spatial resolution, with the added benefit of being able to produce predictions compliant with physical laws. Such \text{dynamical downscaling} approach is however considerably more involved as it require substantial computational and storage resources, as well as considerable expertise to set up WRF properly. As such, our proposed \textit{statistical downscaling} approach is considerably faster and easier to implement without specialized computational resources. The proposed method of adjustment of a simulation via ground observation can also be seen as a bias correction approach, i.e., a method to correct simulations (see, e.g., \cite{yua19, kim15} and \cite{ho12,haw13} for a general review). While a large body of literature in geoscience focuses on bias correction as a means to adjust the first \citep{hemer,lian} and possibly the second moment \citep{teu,li} of the marginal distribution, such approach can be used also to adjust non-Gaussian features, similarly to other recent efforts \citep{pia12,vra14}. 

The proposed statistical model is scalable to future reanalysis data products with even higher spatial resolution, owing to the finite volume approximation of the SPDE generating the spatial model. Even more realistic downscaled patters could be generated if additional physical variables such as temperature and humidity could be considered as covariates. An incorporation of covariates could be performed either as the latent Gaussian model in \eqref{eqn:latent2}, as suggested in this work, or as as additional input of the scalar or vector field which dictate the deformation of the SPDE model. This could be implemented assuming either a linear contribution, or a non-linear one by means of neural networks \citep{hu22}. In principle, multiple variables could be modeled jointly. However, this would considerably increase both the methodological challenge and the computational overhead, as fast, flexible, multivariate and non-Gaussian global models are currently an active area of investigation \citep{gen15}. 

\section*{Acknowledgements}
This research is supported by grant NSF DMS 2014166.

\bibliographystyle{plainnat}

\bibliography{biblio}

\end{document}